\documentstyle[12pt]{article}

\begin{document}
\bibliographystyle{unsrt}
\textwidth 800pt
\large
\begin{center}
\underline{Conditions on the existence of localized excitations in}
\\
\underline{nonlinear discrete systems.  }
\vspace{2cm}\\ \large S. \vspace{0.5cm}Flach$^*$ \\
\normalsize
Department of Physics, Boston University,\\
Boston, Massachusetts 02215, \small flach@buphy.bu.edu
\vspace{1cm}
\\
\end{center}
$^*$ Present address: Max-Planck-Institut f\"ur Physik
Komplexer Systeme, Bayreuther Str. 40, D-01187 Dresden, Germany
\vspace{1.5cm}
\large
\\
\normalsize
ABSTRACT \\
We use recent results that localized excitations in nonlinear
Hamiltonian lattices can be viewed and described as multiple-frequency
excitations. Their dynamics in phase space takes place on tori of
corresponding dimension. For a one-dimensional Hamiltonian
lattice with nearest neighbour interaction we transform the problem
of solving the coupled differential equations of motion into a
certain mapping $M_{l+1}=F(M_l,M_{l-1})$, where $M_l$ for every $l$
(lattice site) is a
function defined on an infinite discrete space of the
same dimension as the torus.
We consider this mapping in the 'tails' of the localized
excitation, i.e. for $l \rightarrow \pm \infty$. For a generic
Hamiltonian lattice the thus linearized mapping is analyzed. We find
conditions of existence of periodic (one-frequency) localized excitations
as well as of multiple frequency excitations.
The symmetries of the solutions are obtained.
As a result we find that the existence of localized excitations
can be a generic property
of nonlinear Hamiltonian lattices in contrast to nonlinear
Hamiltonian fields.
\vspace{0.5cm}
\newline
PACS number(s): 03.20.+i ; 63.20.Pw ; 63.20.Ry
\newline
Date: 05/28/94
\newline
{\sl Physical Review E, accepted}.
\newpage
\section{Introduction}
Localization phenomena are of interest in nearly any branch
of physics. In this paper we will deal
with translationally invariant systems.
In that spirit we deal with
localization phenomena which appear due to {\sl intrinsic}
properties of the underlying system in contrary to
{\sl extrinsic} sources (e.g. defects).
There has been considerable success in the demonstration of
the occurence of localized vibrations in Hamiltonian lattices
\cite{st88},\cite{jbp90},\cite{bkp90},\cite{cp90},\cite{dpw92},\cite{ff93}.
The pure fact of the possibility of vibrational
localization is astonishing because of the translational
invariancy of the underlying lattice (i.e. no defects of any kind
are necessary). The necessary localization condition
has to be the nonlinearity of the lattice, since in the case
of a linear lattice the problem is integrable and only extended
degrees of freedom can be found.

Most of the knowledge about vibrational localization is
restricted to simple one-dimensional lattices, usually
with one degree of freedom per unit cell and nearest neighbour
interaction. These systems belong to the class of
Fermi-Pasta-Ulam (FPU) models (see e.g. \cite{jf92}).
If one adds an external potential (field)
which is periodic with the periodicity of the FPU system, one
enters the world of Klein-Gordon lattices. The important property
of vibrational localization in all these systems is that it is
not of topological origin, i.e. we can consider a system with
only one minimum in the potential energy function (the groundstate)
and will find vibrational localization. If the potential energy function
has several minima one can construct static kink solutions,
i.e. static configurations of the system which link different
groundstates (minima) of the model. Those static kink solutions
can be either minima or saddle points in the potential energy function.
The concept of vibrational localization is
then extendable to those minima (kinks). One simply considers
excitations of the lattice above this (kink) minimum. Because
the static kink himself breaks the translational invariance of
the underlying lattice, a linearization and diagonalization of
the fluctuations around kinks usually yields several eigenmodes
which correspond to localized vibrations
\cite{dhn74},\cite{gj75},\cite{ws78},\cite{bw89}.
The rest of the eigenmodes
are deformed phonon-like extended degrees of freedom. Here we find
a difference with the nontopological vibrational localization -
namely localization in the linear limit. Of course it becomes only
possible because the static kink already broke the translational symmetry.
It is worthwhile to notice that the topological-induced
vibrational localization is comparable to the well-known vibrational
localization in a linear lattice with defects \cite{hb83}.

Recently we were able to achieve progress in the understanding
of non-topological vibrational localization
\cite{fw3},\cite{fw2}. First we noticed
that it is possible to find nonlinear localized excitations (NLEs)
which are essentially located on a very few particles. By analyzing the
dynamics of the lattice {\sl on finite time scales}
we found that in general the NLE is a
many-frequency excitation, where the number of frequencies $n$ is
equal to the number of particles which are essentially involved in
the excitation. By studying a reduced system which is given by the dynamics
of the part of the lattice where the NLE is located (with proper
defined boundary conditions) we showed that the NLEs in the
original lattice correspond to regular motions of the reduced system
on $n$-dimensional tori in its phase space. Moreover the NLE solution
of the original lattice evolves on nearly the same torus. Thus
it becomes possible to systematically study the NLE properties
by checking the phase space structure of the reduced system. As
a result we found that chaotic motion in the reduced system does not
yield NLEs, as well as certain regular islands in its phase space
which are well separated (by separatrices) from the NLE-regular
islands. We were also able to attack the problem of movability of NLEs
by considering a certain separatrix in the phase space \cite{fw6}.
Still the question remains, whether NLEs can be {\sl exact } solutions
of the lattice equations of motion.

Much less is known about more complicated models. First we notice
that the above cited approach in the simplest one-dimensional
systems is easily extended to higher dimensions or to
more than one degrees of freedom per unit cell. Indeed recent
reports confirm the existence of NLEs in diatomic chains
\cite{ma92},\cite{bks93},\cite{ats93} as
well as in two-dimensional FPU systems
\cite{bkp90},\cite{ff93}. A systematic analysis
of the NLE properties in two-dimensional Klein-Gordon lattices
is carried out in \cite{fw8}. Thus there is no doubt that by studying
one-dimensional non-topological NLEs we are not restricting ourselves
to exotic cases. The more complex reality will be covered.
This is fundamentally different from the topological kink solutions.
The reason is that the NLE solutions are not topologically induced.
The only source of their existence is the nonlinearity of the
underlying lattice.

All presently known theoretical approaches to describe NLEs
make the assumption that the NLEs exist. Then one can proceed
in the description of their properties. In this paper we
want to present an approach to the problem
of the existence of NLEs. We will study the simplest one-dimensional
cases. As we have shown, this restriction is of no fundamental
significance. We will prove several conditions when NLEs
can not exist. Thus the remaining cases are the ones one has
to choose if (possibly) NLEs exist. We will use the knowledge
about the interpetation of NLEs in terms of actions and angles
and consider a general ansatz. Then we reduce the problem of
solving coupled ordinary differential equations to a (still
highly complicated) set of coupled algebraic equations. The variables
are certain Fourier-components. We will consider these algebraic
equations as a mapping of a function defined on a $n$-dimensional
lattice. By the definition of an NLE we show that the algebraic
equations decouple in the tails of the supposed existing NLE solution. Then we
analyse the decoupled mapping in the tails and calculate the
eigenvalues of the linear map. We observe cases when periodic
NLEs ($n=1$) can not exist. We prove that strictly speaking there
exist no solutions for $n \geq 2$, which implies that many-frequency
NLEs are unstable.
Still this bare fact allows for no strict conclusion about the
typical decay time of many-frequency NLEs.
We relate our studies to previous work on stability analysis
of NLEs.

\section{Formulation of the problem}
We will consider the dynamics of a simple one-dimensional
Hamiltonian lattice with the following Hamiltonian:
\begin{equation}
H=\sum_l \left(\frac{1}{2} P_l^2 + V(X_l) + \Phi(X_l - X_{l-1})\right) \;\;\;.
\label{1}
\end{equation}
Here $P_l$ and $X_l$ are canonically conjugated momentum and displacement
of the $l$-th particle, $V(z)$ and $\Phi(z)$ are the potentials
of the external field and nearest neighbour interaction respectively.
We do not consider incommensurabilities between these two potentials,
thus we assume that there exists at least one groundstate of \ref{1}
(minimum of the potential energy) such that without loss of generality
$X_l=0$ for this groundstate. We specify the potential terms
in \ref{1} in form of an expansion around this groundstate:
\begin{eqnarray}
V(z)= \sum_{k=2}^{\infty} \frac{1}{k!}v_k z^k \;\;, \label{2} \\
\Phi(z) = \sum_{k=2}^{\infty} \frac{1}{k!}\phi_k z^k \;\;. \label{3}
\end{eqnarray}
The NLE solution is in its general form assumed to be given by the motion of
the phase space trajectory of \ref{1} on a $n$-dimensional torus
\cite{fw3},\cite{fw2}.
Consequently the solution has to have the form:
\begin{equation}
X_l(t)=\sum_{k_1,k_2,...,k_n=-\infty}^{+\infty}
f_{lk_1k_2...k_n}e^{i(k_1\omega_1+k_2\omega_2+...+k_n\omega_n)t}\;\;.
\label{4}
\end{equation}
The localization property of \ref{4} is defined by the boundary
condition
\begin{equation}
f_{lk_1k_2...k_n}\mid_{l \rightarrow \pm \infty} \rightarrow 0 \;\;\;.
\label{4.5}
\end{equation}
Here we are excluding from the definition of an NLE
localized pulses on carrier waves where the carrier wave doesn't
decay far away from the center of the pulse.
Since \ref{4} is by assumption a solution of the equations of motion
\begin{equation}
\ddot{X_l}=\dot{P_l}=-\frac{\partial H}{\partial X_l}\;\;\;, \label{5}
\end{equation}
we can insert \ref{4} into \ref{5} and try to solve for the Fourier
coefficients on the right-hand side of \ref{4}. Using
\begin{eqnarray}
\ddot{X}_l(t)=-\sum_{k_1,k_2,...,k_n=-\infty}^{+\infty}
y_{\vec{k}}(\vec{\omega})
f_{lk_1k_2...k_n}e^{i(k_1\omega_1+k_2\omega_2+...+k_n\omega_n)t}\;\;,
\label{6} \\
y_{\vec{k}}(\vec{\omega}) =
(k_1\omega_1+k_2\omega_2+...+k_n\omega_n)^2\;\;\;, \label{7} \\
\vec{k}=(k_1,k_2,...,k_n)\;\;,\;\;\vec{\omega}=
(\omega_1,\omega_2,...,\omega_n) \;\;, \label{7.5}
\end{eqnarray}
we get Fourier series on the left and right hand sides of
\ref{5}. The only possiblity of satisfying the obtained equation
is to collect terms with equal exponents on both sides and to
put the prefactors equal to each other. Then we obtain a highly
complicated coupled set of algebraic equations for the Fourier
coefficients and the frequencies.
Because we consider nearest neighbour interaction \ref{1}
we can formally write down the resulting set of equations:
\begin{equation}
M_{l+1,\vec{k}}=F(\{ M_{l,\vec{k'}} \} ,
\{ M_{l-1,\vec{k''}} \} ) \;\;\;. \label{8}
\end{equation}
Here we introduced a function $M_{l,\vec{k}}$ which is defined on
a discrete $n$-dimensional lattice. The lattice is given by
all combinations of $\{k_1,k_2,...,k_n\}$ where each integer $k_{n'}$
varies from $-\infty$ to $+\infty$. We have
\begin{equation}
M_{l,\vec{k}}=f_{lk_1k_2...k_n}\;\;\;. \label{9}
\end{equation}
The mapping formally derived in \ref{8} reminds us of the well-known
two-dimensional mappings which were used in order to
study static kink properties, commensurate-incommensurate
transitions and breaking of analyticity \cite{sa93}. We can view \ref{8}
as a first step of implementing the fruitful ideas for static
topologically induced structures in the dynamical problems
of nontopological NLEs in Hamiltonian lattices.

Let us study \ref{8} in the tails of the NLE, i.e. for
$l \rightarrow \pm \infty$ where \ref{4.5} holds by assumption. In the
generic case $v_2$ and $\phi_2$ from \ref{2},\ref{3} will be
nonzero. Then we can write down the mapping \ref{8} explicitely.
We will do it without loss of generality for $l \rightarrow +\infty$.
The corresponding formula for large negative $l$ can be obtained
by substituting $l'=-l$. We find
\begin{equation}
M_{l+1,\vec{k}}=(\kappa_{\vec{k}}(\vec{\omega})+2)
M_{l,\vec{k}}-M_{l-1,\vec{k}}
\;\;\;. \label{10}
\end{equation}
Here we have introduced another function on the $n$-dimensional
discrete space which is given by
\begin{equation}
\kappa_{\vec{k}}(\vec{\omega})=\frac{v_2-y_{\vec{k}}(\vec{\omega})}
{\phi_2}\;\;\;. \label{11}
\end{equation}
Equation \ref{10} is linear and thus every component of $M$ in the
$n$-dimensional discrete space decouples in this equation from
all other components. Introducing
\begin{equation}
G_{l,\vec{k}}=M_{l,\vec{k}}-M_{l-1,\vec{k}} \;\;\;\label{12}
\end{equation}
we finally arrive at a two-dimensional mapping for every component of
$M_{l,\vec{k}}$ which reads
\begin{eqnarray}
M_{l+1,\vec{k}} & = & \kappa_{\vec{k}}(\vec{\omega})
M_{l,\vec{k}} + G_{l,\vec{k}} + M_{l,\vec{k}}\;\;\;, \label{13} \\
G_{l+1,\vec{k}} & = & \kappa_{\vec{k}}(\vec{\omega})
M_{l,\vec{k}} + G_{l,\vec{k}} \;\;\; . \label{14}
\end{eqnarray}
Let us stress that under the assumption of an existing NLE solution
the linearization of the map in the tails of the NLE is arbitrarily
correct, if the distance from the NLE center is large enough.
This mapping has a fixed point for $M_{l,\vec{k}}=0$ and $G_{l,\vec{k}}=0$.
It is characterized by the matrix $A$:
\begin{equation}
A =
\left(
\begin{array}{lr}
1 & \kappa_{\vec{k}}(\vec{\omega})  \\
1 & 1+\kappa_{\vec{k}}(\vec{\omega})
\end{array}
\right)\;\;\;. \label{matrix}
\end{equation}
We have
\begin{equation}
{\rm det} (A)=1 \;\;\;. \label{16}
\end{equation}
Thus the mapping \ref{13},\ref{14} is symplectic and volume preserving.
For the eigenvalues of $A$ we find
\begin{eqnarray}
\lambda_{\pm} & = & 1+\frac{\kappa_{\vec{k}}(\vec{\omega})}
{2}\pm \sqrt{(1+\frac{\kappa_{\vec{k}}(\vec{\omega})}{2})^2 - 1}\;\;,
\label{17} \\
\lambda_+ \lambda_- &  =  & 1 \;\;. \label{18}
\end{eqnarray}
We can consider three cases:
\begin{equation}
a)\;\;\kappa_{\vec{k}}(\vec{\omega}) > 0 \;\;: \;\;\;
0 < \lambda_- < 1 \label{19}
\end{equation}
i.e. $\lambda_-$ is real. Especially $\lambda_-(\kappa_{\vec{k}}(\vec{\omega})
\rightarrow 0
)\rightarrow 1$ and $\lambda_-(\kappa_{\vec{k}}(\vec{\omega})
\rightarrow \infty)
\rightarrow 0$.
\begin{equation}
b) \;\; \kappa_{\vec{k}}(\vec{\omega}) < -4 \;\;:
-1 < \lambda_+ < 0 \label{20}
\end{equation}
i.e. $\lambda_+$ is real. Especially $\lambda_+(\kappa_{\vec{k}}(\vec{\omega})
\rightarrow -
\infty) \rightarrow 0$ and $\lambda_+(\kappa_{\vec{k}}(\vec{\omega})
\rightarrow -4)
\rightarrow -1$.
\begin{equation}
c) \;\;-4 \leq \kappa_{\vec{k}}(\vec{\omega}) \leq 0 \;\;:
\;\;\; \left|\lambda_+\right|
= \left|\lambda_-\right| = 1 \label{21}
\end{equation}
i.e. $\lambda_{\pm}$ are complex conjugated numbers on the unit circle.
Consequently in cases a) and b) the fixed point of the
mapping is a saddle point, i.e. there exists exactly one direction
(eigenvector) in which the fixed point can be asymptotically reached
after an infinite number of steps. In case c) the fixed point is
a marginally stable elliptic point, i.e. starting from any direction
the fixed point can be never reached after an infinite number of steps,
instead the mapping will produce a (deformed) circle around the
fixed point. Thus we find that case c) (\ref{21}) contradicts the
localization condition \ref{4.5}.

\section{Single-frequency localized excitations}

Let us consider $n=1$. Then the NLE solution is periodic
(cf. \ref{4}). Equation \ref{11} can be simplified to
\begin{equation}
\kappa_{\vec{k}}(\vec{\omega})=\frac{v_2-k_1^2\omega_1^2}{\phi_2}
\;\;\;. \label{24}
\end{equation}
The frequencies $\omega_q$ for small-amplitude phonons around
the considered groundstate of \ref{1} (where $q$ is the wave number)
are related to the parameters $v_2$ and $\phi_2$ by
\begin{equation}
v_2 \leq \omega_q^2 \leq v_2 + 4 \phi_2 \;\;\;. \label{25}
\end{equation}
Then it follows that case c) given in \ref{21} is identical with
\begin{equation}
c) \;\;\; k_1^2\omega_1^2 = \omega_q^2 \;\;\;. \label{26}
\end{equation}
We find that a single frequency NLE (periodic localized solution)
can not exist if any multiple of its fundamental frequency
equals any phonon frequency. The reason is that we can not satisfy
\ref{26} and \ref{4.5} simultaneously because of \ref{21}.

In \cite{fw3},\cite{fw2} we have shown that under the assumption of the
existence of a single frequency NLE its stability with respect
to small amplitude phonon perturbations will depend on the
fundamental NLE frequency. We found that if $\omega_q/\omega_1=n/2$,
$n=0,1,2,...$ then the small perturbation will grow. Consequently the
NLE would collapse. Here now we find that if $n=2k_1$ (even $n$)
then the NLE itself could not exist. If $n=2k_1+1$ (odd $n$)
then the NLE could exist, but would be unstable against phonon
perturbations.

One can interprete \ref{26} as a definition of nonexistence bands on
the $\omega_1$-frequency axis for different $k_1$. Introducing a
normalized frequency $\tilde{\omega}_1=\omega_1/\sqrt{v_2}$
and normalized interaction $\tilde{\phi}_2=\phi_2/v_2$
(only if $v_2 \neq 0$) those bands are given by
\begin{equation}
\frac{1}{k_1^2} \leq \tilde{\omega}_1^2 \leq \frac{1+4\tilde{\phi}_2}{k_1^2}
\;\;\;. \label{27}
\end{equation}
For any finite $\tilde{\phi}_2$ and small enough $\tilde{\omega}_1$
the nonexistence bands \ref{27} will start to overlap. Consequently
there will be always a lower bound on allowed NLE frequencies. Still
some existence windows are possible in the phonon gap ($\tilde{\omega}_1
< 1$). However with increasing $\tilde{\phi}_2$  more nonexistence
bands will overlap and at the critical value $\tilde{\phi}_2^c=3/4$
all nonexistence bands \ref{27} overlap. For all values of
$\tilde{\phi}_2 > 0$ and any value of $\tilde{\omega}_1 > 1+4\tilde{\phi}_2$
condition \ref{27} is not satisfied. That means that independent
of the values of the parameters $v_2 \geq 0$ and $\phi_2 > 0$ periodic
NLE solutions are allowed with frequencies above the phonon band.
{}From the above results it follows that model \ref{1} always allows
for periodic NLEs with frequencies above the phonon band. But if
the model has a nonvanishing lower phonon band edge ($v_2 > 0$)
then periodic NLEs with frequencies in the phonon gap are allowed
if the phonon band width is small enough compared to the lower
phonon band edge.

Now let us make some statements about symmetries of periodic NLEs
if they exist. If the frequency of the NLE is above the phonon band
then it follows $\kappa_{\vec{k}}(\vec{\omega}) < -4 $ for all $k_1$.
This corresponds to case b) in \ref{20}. Then we have $-1 < \lambda_+
< 0$. Consequently $-1 < M_{l+1,\vec{k}}/M_{l,\vec{k}} < 0$
for all $l$ in the tail of the NLE. Thus we find a coherent out-of-phase
type of the motion of neighbouring particles in the tails of the NLE solution
because of \ref{4}: $-1 < X_{l+1}(t)/X_l(t) \leq 0$ if defined.
If the frequency of the periodic NLE is in the phonon gap ($v_2 > 0$)
things become more complicated. Namely there will always exist a certain
finite integer $k_c$ such that for $k_1 < k_c$ it follows
$\kappa_{\vec{k}}(\vec{
\omega})> 0$ which corresponds to case a) in \ref{19}.
The corresponding Fourier components \ref{9} would yield in-phase type
of motion in the tails of the NLE solution. However for all $k_1>k_c$
the case b) in \ref{20} applies. Those Fourier components would
yield out-of-phase motion. Numerical findings indicate that usually
the Fourier components decay very fast with increasing $k_1$
\cite{dpw92}.
Then we could expect overall in-phase type of motion. However if
the frequency $\omega_1$ becomes smaller then it is well-known that
the decay in the Fourier components with increasing $k_1$ slows
down. Thus we have to expect a complicated mixture of in- and out-of-phase
type of motion.

\section{Many-frequency localized excitations}

Let us consider $n=2$. Then
in analogy to \ref{26} case c) in \ref{21} applies if
\begin{equation}
v_2 \leq (k_1\omega_1+k_2\omega_2)^2 \leq v_2+4\phi_2
\;\;\;. \label{28}
\end{equation}
Now it is possible to show that there exists an infinite number
of pairs of the integers $(k_1,k_2)$ such that \ref{28}
is satisfied if the ratio $\omega_1/\omega_2$ is irrational
and $v_2 \geq 0$ and $v_2 > 0$ (cf. Appendix).
Thus strictly speaking there exist no exact two-frequency
NLEs. The proof for $n \geq 3$ is then straightforward and
yields the same result. Of course this fact does not tell anything
about decay times of many-frequency NLEs. It only states that
many-frequency NLEs can not exist for infinite times.
It seems to be logical to assume that the decay times are sensitive
to the pair of the lowest integers $(k_1,k_2)$ for which
\ref{28} holds in the case $n=2$. Indeed numerical simulations
\cite{fw3},\cite{fw2}
show extremely weak decay of two-frequency NLEs in Klein-Gordon
chains, i.e. the characteristic decay time is several orders
of magnitude larger than internal oscillation times.

\section{Time-space separability}

Recently there were reports in the literature
where for systems of type \ref{1} periodic NLE solutions
with a property of time-space separability were proposed to exist
\cite{ysk93}.
In more detail this property implies the existence of
a master function $G(t)$ in time such that the NLE solution
can be given by
\begin{equation}
X_l(t)=A_l G(t) \;\;, \;\; A_l\mid_{l \rightarrow \pm \infty}
\rightarrow 0 \;\;. \label{5-1}
\end{equation}
Without loss of generality one can set ${\rm max}(A_l)=1$. In terms of
the introduced Fourier components in \ref{4} ansatz \ref{5-1}
imposes a rather strong symmetry on the Fourier components - namely
they have to be equal to each other at different lattice sites up
to a universal scaling number. If we insert \ref{5-1} into the
equations of motion \ref{5} we find
\begin{equation}
\ddot{G}(t) = - \sum_{k=2}^{+\infty}\eta_kG^{k-1}(t) \;\;\;.\label{5-2}
\end{equation}
With \ref{2},\ref{3} we can specify the constants $\eta_k$:
\begin{equation}
\eta_k = \frac{1}{A_l}[v_kA_l^{k-1}+\phi_k\{ (A_l - A_{l-1})^{k-1}
- (A_{l+1}-A_l)^{k-1}    \}]\;\;. \label{5-3}
\end{equation}
If $\eta_k$ in \ref{5-3} would depend on $l$ then the differential
equation for the master function $G(t)$ in \ref{5-2} would yield different
solutions for different lattice sites (cf. e.g. \cite{via73}). Then
we contradict the original ansatz \ref{5-1}. Thus we have to assume
that $\eta_k$ in \ref{5-3} is independent of $l$. Hence we generate a
set of two-dimensional mappings for the vector $(A_l,A_{l-1})$ in
\ref{5-3} for different $k$.

Let us show that generically we can not satisfy all those mappings.
For that we consider a Klein-Gordon lattice, i.e. $\phi_k=0$ for
$k \geq 3$, at least one $v_k$ is nonzero for $k \geq 3$. Then
we consider \ref{5-3} for that specific $k$:
\begin{equation}
\eta_k = v_k A_l^{k-2}\;\;,\;\; k\geq 3\;\;. \label{5-4}
\end{equation}
{}From \ref{5-4} it would follow that $A_l=1$ for all $l$ in contradiction
to \ref{5-1}. Thus we would have to conclude that either no NLEs
exist in Klein-Gordon lattices or that they do not obey separability
property \ref{5-1}. The existence of periodic NLEs in various
Klein-Gordon systems is verified numerically with very high
probability ( i.e. currently periodic NLEs can be generated without
any measurable energy loss over $10^6$ time periods of the internal
NLE oscillation, cf. Section 6). Consequently the
ansatz \ref{5-1} is restricted to a certain subclass of systems which
excludes the Klein-Gordon systems. Now let us consider any system
such that for only (or at least) one integer $k$ we have $v_k=\phi_k=0$.
Let us assume that we have periodic NLE solutions which satisfy
\ref{5-1}. Perturbing this system with any small but finite
$v_k \neq 0$ we can again consider the mapping for that $k$ and yield
\ref{5-4}. Consequently no NLE solution would be allowed.
Thus the separability ansatz
\ref{5-1} becomes a real exotic property.

There is one
case when we can at least expect that \ref{5-1} holds.
Namely when we have to satisfy only one mapping of the type \ref{5-3}.
This can happen e.g. if $v_k=0$ for all $k$ and $\phi_{k_0} \neq 0$ for
$k_0 \leq 4$ and $k_0$ even and $\phi_k = 0$ for all $k \neq k_0$.
One can prove that if \ref{5-3} is rewritten for that specific case
in terms of a mapping of the two-component vector $(A_l,A_{l}-A_{l-1})$
then the mapping will be nonconservative (it does not preserve
phase volume). A numerical test for $k_0=4$ yields that a periodic
NLE solution can be found for $\eta_{k_0}=9.5843773776314...$.
The amplitudes for that solution read $...,0.004795,-0.1657879,1,-1,
0.1657879,-0.004795,...$.
First we notice that indeed for such special systems it can be
possible to generate a NLE solution with time-space separability property
\ref{5-1}.
However our result also indicates that
even for such a highly nonlinear model (no linear dispersion)
the NLE if it exists has no compact structure, i.e.
the amplitudes $A_l$ are not exactly zero outside a finite
volume of the solution. That fact can also be observed by looking
at \ref{5-3}. If outside of a finite volume of a solution all
amplitudes would be equal to zero, by inverting the mapping we
would generate zeros for all amplitudes in the excluded finite volume
thus contradicting the ansatz that at least one amplitude is equal to 1.
As a consequence the claimed compacton structure of a periodic
NLE solution in \cite{ysk93} is wrong and based on a simple calculation
error.

\section{The decay of single-frequency NLEs in the tails}

Let us consider an existing periodic NLE for system \ref{1}
with $v_2> 0$, $\phi_2 > 0$. Then in the tail of the NLE
(without restriction for large positive $l$) every Fourier
component obeys a two-dimensional mapping given in \ref{13},\ref{14}.
The eigenvalue describing the decay is given by \ref{17},\ref{24}.
Then the decay of every Fourier component will be governed by
an exponential law $\sim ({\rm sgn}(\lambda))^l{\rm e}^{{\rm ln}\mid
\lambda \mid l}$.
The absolute value of $\lambda$ will depend on the number $k$ of the
Fourier component. For $l \rightarrow \infty$ only the Fourier
component with $\mid \lambda \mid$ closest to one will
be present, i.e. all other Fourier components will decrease
exponentially fast compared to the remaining one. The answer to
the question which Fourier component 'survives' depends on the
frequency $\omega_1$ and on the parameters $v_2$ and $\phi_2$.
Only if $\omega_1$ is above the phonon band then can we state
that the 'surviving' Fourier component is that with $k=1$.

In the following we will formulate two predictions for
single-frequency NLEs in order to test them in a given
realization. The first prediction was already formulated
above - if an assumed periodic NLE (characterized by its fundamental
frequency $\omega_1$) exists, the amplitude decay in the tails
of the NLE (i.e. where the nonlinear terms in the equations of motion
can be neglected) is governed by an exponential law with
exponent $ln|\lambda |$. Let us consider the case of a $\phi^4$
chain, i.e. $V(z)=z^2-z^3+0.25 z^4$ and $\Phi(z)=0.5 C z^2$.
Consequently we have $v_2=2$, $\phi_2=C$.
Periodic NLEs were investigated in \cite{fw3},\cite{fw2} on finite
time scales for the case $C=0.1$. We use a particular realization
with energy $E=0.256$ (for details on the numerics the
reader should consult the
original work \cite{fw3},\cite{fw2}). The initial conditions
on the lattice (3000 particles) are found from analyzing the
elliptic fixed point of a reduced problem \cite{fw2}.
The frequency of the periodic localized object is found to
be $\omega_1=1.177$ (cf. Fig.10 in \cite{fw2}).
After
a waiting time of $T=3 \cdot 10^5$ the energy stored in five
particles around the excitation center is still $E(T)=0.256$.
There is absolutely no drift (radiation) observable.
The symmetric  amplitude distribution in this assumed NLE solution is
shown in Fig.1 in a semilogarithmic plot (open squares).
An evaluation of the eigenvalues corresponding to the different
Fourier components (cf. \ref{17}-\ref{24}) reveals in this
case that the eigenvalue with absolute value closest to one is
given for $k=1$ and reads $\lambda=0.12465$ ($\kappa_1 = 6.1467)$.
Consequently in the tail of the solution (where the nonlinear
terms in the equations of motion can be neglected) we find an
exponential amplitude decay with exponent ${\rm log}_{10}|\lambda|=-0.904$.
The dashed line is this prediction of the exponential decay.
Note that in the semilogarithmic plot in Fig.1 this decay law
appears as a straight line. This line fits the measured
amplitude decay down to amplitudes of $10^{-3}$. The deviations
for smaller amplitudes (farther away in the tail) are due to
the fact, that the numerical realization of the NLE solution is
always accompanied by the existence of phonons. The phonons
with nearly zero group velocities (wave numbers close
to band edges) are practically not moving. These small phonon
contributions increase (additively, as has to be expected
in a linearizable tail of a NLE) the amplitudes of the particles
in the NLE solution. We have no knowledge of any result
in the literature, where from the measurement of the frequency of
an assumed NLE solution the amplitude decay in the NLE tails was
successfully predicted.

The second prediction we wish to formulate is, that if the
NLE frequency $\omega_1$ is restricted to be in the nonzero
gap of the phonon spectrum, then there exists a nonzero
value of $\omega^{(m)}_1$ such that the exponential decay in the
NLE tails will be weaker for all other frequencies $\omega_1$
(still belonging to the gap). Let us explain why this statement
follows out of the previous considerations.
First the frequency $\omega_1$ has to be larger than the
phonon band width - else one (or more) of multiples of $\omega_1$
will always lie in the phonon band. Secondly if $\omega_1$
is slightly below the lower phonon band edge, then the decay
will be very weak. Lowering the $\omega_1$ we increase the decay exponent,
but since $2\omega_1$ is coming closer to the upper phonon band edge, there
will be a certain frequency when the decay in the tails will be governed
by the second harmonic rather than the first harmonic of $\omega_1$.
Let us calculate $min(1-|\lambda(\kappa_k)|)$ with $|\lambda(\kappa_k)| \leq 1$
and the parameters of the above introduced $\phi^4$ chain for
different $\omega_1$. The result is shown in Fig.2. We see indeed
that there is a maximum in the decay exponent in this given example
for $\omega_1^{(m)}=0.938$. As it can be observed from Fig.10 in \cite{fw2},
there are two NLE energies $E_1=0.336$ and $E_2=0.85$ which correspond
to this particular frequency $\omega_1^{(m)}$. The prediction is thus,
that by varying the energy of the periodic NLEs, in a simulation
one should observe maximum amplitude localization at these two
energies $E_1,E_2$. From the numerical runs reported in \cite{fw2}
we can calculate a normalized amplitude entropy $\sigma_a$ which
measures the amplitude distribution of the solution.
It is defined in analogy to the energy entropy which is defined in \cite{fw2}:
$\sigma_a = -1/({\rm ln}(N)) \sum_l a_l{\rm ln} a_l$
with $a_l = A_l/\sum_l A_l$.
Here $A_l$ is the amplitude of particle $l$.
{}From the definition we have $0 < \sigma_a < 1$.
Maximum localization corresponds to a minimum of the normalized entropy.
$\sigma_a$ as a function of the NLE energy is shown
in Fig.3 (filled triangles and solid line). The predicted positions
of the minima of $\sigma_a(E)$ at $E_1,E_2$ are given by vertical
dashed lines. The predicted positions of the minima agree with
the measured ones within the energy grid of the numerical experiment.

In order to demonstrate the usefulness of the above considered
spatial decay laws, we have implemented a numerical method
in order to solve the {\sl full nonlinear} equations for the
Fourier coefficients \ref{8} for a periodic NLE ($n=1$) on a finite
lattice. We have choosen $v_2=1$, $v_4=-1$, $v_{k\neq 2,4}=0$,
$\phi_2=0.1$, $\phi_{k\neq 2} =0$ and a system size of 100 particles
with periodic boundary conditions. The NLE frequency was choosen
to be below the phonon band: $\omega_1=0.8$. The resulting Fourier
coefficients were computed on every lattice site with a maximum
multiple of the fundamental frequency $|k|_{max}=30$ and a numerical
error of $10^{-20}$. The result is shown in Fig.4 for thirty lattice
sites around the NLE center in a semilogarithmic plot. Clearly a
$k$-dependent exponential decay of every Fourier component is found.
Moreover in Fig.5 we plot the measured exponents (slopes) as found from
the numerical solution and compare them to the theoretical prediction.
The agreement is very good.

What happens if $v_2=0$ and $\phi_2=0$? We can give a particular
answer for the special case $\phi_4 \neq 0$ and all other
coefficients in \ref{2},\ref{3} being zero. This case was already
discussed in the preceding section. If we know that a periodic
NLE exists we can consider the mapping \ref{5-3}. We linearize
the map near the saddle fixed point and find
in leading order a decay of the amplitudes according to the
law ${\rm e}^{-a \exp{bl}}$ where $a$ and $b$ are positive numbers
which depend on the parameters of the model Hamiltonian.

\section{Conclusion}

Nonlinear localized excitations (NLEs) might exist in nonlinear
lattices for infinite times (i.e. they can be exact solutions
of the equations of motion) if they are periodic in time and
all multiples of the fundamental frequency are outside the phonon
band. This is only possible because of the discreteness of the
underlying lattice. A continuum system would have no upper phonon
band edge, thus resonance would always be possible. Exceptions are
classes of nongeneric systems where the resonance condition holds but
the coupling between the NLE and the phonons is exactly zero.

Let us note that we have not shown that periodic NLEs {\sl will} exist
on a given nonlinear lattice. We have restricted the possibilities
of NLE solutions to time-periodic NLEs.

If the NLE solution is described by two or more fundamental frequencies
then there will be always resonance with phonons and thus
those solutions can not exist for infinite times, i.e. they are not
exact solutions of the equations of motion. Nothing can be said up to now
about
the lifetimes of such solutions. Numerical test show that lifetimes
can be very large compared to internal oscillation times.
If one considers nongeneric lattices without linear dispersion (phonons)
then many-frequency NLEs might be exact solutions of the equations
of motion.

Space-time separability in the NLE solution can appear only in very nongeneric
cases.
The construction of mappings for Fourier coefficients in
higher dimensional Hamiltonian lattices seems to be more complicated
but this is a technical question. The
construction of the existence conditions for NLEs indicate that one
can generalize them for
higher dimensions too.

Let us make some comments on the results rpesented in this paper. First we
have assumed that the NLE solution (if it exists) is given
by a regular motion on a torus in the phase space of the system
(discrete spectrum).
It is very hard to believe that NLE solutions can exist if
the corresponding orbit belongs to a stochastic part of the
system's phase space (continuous spectrum).
Assuming the NLE solutions
we search for are regular, all the subsequent steps in the presented analysis
are free from any simplifications, approximations or conjectures.
We have formulated the leading order amplitude decay of an assumed
NLE solution in its tails. We were thus able to formulate two
nontrivial predictions and test them by comparing to numerical
experiments.

Recently it was shown by using the properties of map \ref{8},
that the existence of NLE solutions is equivalent to finding
a common point of two separatrix manifolds of the nonlinear map
\cite{fw9}.
It was argued that the corresponding task is not overdetermined
and should lead to a discrete set of solutions. Moreover in the
case of an anharmonic Fermi-Pasta-Ulam (FPU) chain with
homogenious potential $v_k=0$, $\phi_{2m}=1$, $\phi_{k\neq 2m}=0$ (where $m$
is any positive integer with $m \geq 2$) the existence of time-periodic
NLEs could be strictly proved \cite{fw9}.

Since the presented analysis was carried out in the tails of
the NLE solution (where the corresponding mappings can be
linearized) the legitimate question arrises, why we do not
have NLE solutions in a purely harmonic lattice. The reason
for the nonexistence of a localized solution in such a
linear lattice is that the two separatrix manifolds of
the saddle fixed point never cross, because the eigenvectors
of the fixed point uniquely define the positions of the
two separatrix manifolds. In a nonlinear system, inverting
the discrete map (i.e. starting in the tail of an assumed existing
NLE and iterating towards the center of the NLE) the increase
of the amplitudes (or Fourier coefficients) will increase the
contributions coming from the nonlinearity. These nonlinearities
will be the reason for the crossing of the separatrix manifolds.
We wish to emphasize that still these arguments can not
be considered as a proof of the existence of NLEs. They provide
the possible source of the NLE existence in the framework of
the presented approach. The aim of the present work was to
analyze the NLE properties in the tails assuming the
NLE existence. The obtained results restrict possible NLE solutions
to time-periodic solutions.

Recently MacKay and Aubry have carried out a proof of existence
of time-periodic NLEs for Hamiltonian networks of weakly coupled
oscillators \cite{ka94}. This strict proof remarkably incorporates Hamiltonian
lattices with and without disorder. Consequently one can see, that
(at least in the limit of weak coupling) the NLEs in a nonlinear
lattice are continuously connected with localized modes of say a
harmonic lattice with defects. These results should once and forever
make it clear, that a harmonic lattice without defects is a very
isolated system, and one should never use it as a reference system
in order to judge results of nonlinear theories.
In the present work the proof of MacKay and Aubry corresponds to the
case $v_{k} \neq 0$, $\phi_k \ll v_2$. According to \cite{ka94}
the spatial decay of NLEs is at least exponential, which can be
specified to be strictly exponential in the present work
without the requirement of weak coupling.
Together with the strict proof of
NLE existence in the mentioned anharmonic FPU chain \cite{fw9}
we thus have an analytical basis of the existence of NLEs in
a broad class of systems.

We were able to construct a mapping for the NLE solutions because
of our knowledge about their proposed dynamical behaviour. This knowledge
comes from the interpretation of the NLE solutions as motion
of phase space points on tori in the phase space. The resulting mapping
is an algebraic problem and can be considered as a task
for itself. Here we meet numerous results for approximate
NLE solutions which use the Rotating Wave Approximation (RWA)
\cite{st88},\cite{th91},\cite{st92},\cite{dpw92},\cite{ats93}.
Within the RWA all Fourier components higher than a certain
hand-choosen order are set to zero. The resulting algebraic
equations can be either solved selfconsistently or by means of
useful iteration procedures.
Only periodic NLEs were considered up to now.

It is interesting to note that recently a combination of the
lowest RWA order (taking into account only the lowest Fourier
component) was reformulated into a mapping
\cite{sw93}. It would correspond to the mapping for the amplitudes
of the NLE solution in the present work under the assumption
of time-space separability.

The fact that the existence of NLEs seem to be a generic property of
(nonlinear) Hamiltonian
lattices implies that either experimental realizations of NLEs
can be found with probability 1 or that the considered model classes
are useless in describing reality. This promises some intriguing
questions for the future.
\\
\\
\\
Acknowledgements
\\
\\
This work would be impossible without the continuous
stimulating interaction with C. R. Willis.
I thank R. S. MacKay and S. Aubry for sending a preprint prior publication,
R. S. MacKay and B. Birnir for interesting discussions and
M. Raykin for helpful hints.
This work was supported
by the Deutsche Forschungsgemeinschaft (Fl200/1-1).
\newpage
\appendix

\section{}

Let us prove that if $a$ and $b$ are two real numbers such
that
\begin{equation}
0 \leq a < b \label{a1}
\end{equation}
and $\omega_1$ and $\omega_2$ are two real numbers such
that
\begin{equation}
0 < \omega_1 < \omega_2 \label{a2}
\end{equation}
and the ratio $\omega_1/\omega_2$ is irrational,
then there always exists an infinite number of pairs of
integers $(k_1,k_2)$ such that
\begin{equation}
a \leq \mid k_1\omega_1 + k_2\omega_2 \mid \leq b \;\;\;. \label{a3}
\end{equation}
Let us introduce
\begin{equation}
\tilde{a}=\frac{a}{\omega_2}\;\;,\;\;\tilde{b}=\frac{b}{\omega_2}\;\;,
\;\; \xi=\frac{\omega_1}{\omega_2}\;\;\;. \label{a4}
\end{equation}
Then \ref{a3} is aquivalent to
\begin{equation}
\tilde{a} \leq \mid k_1\xi + k_2 \mid \leq \tilde{b}\;\;\;. \label{a5}
\end{equation}
Let us choose the numbers $\tilde{c}$ and $\tilde{d}$ such that
\begin{equation}
\tilde{a} < \tilde{c} < \tilde{b}\;\;,\;\;
\tilde{d} = {\rm min}((\tilde{c}-\tilde{a}),(\tilde{b}-\tilde{c}))\;\;\;.
\label{a6}
\end{equation}
We can consider an arbitrary integer $N$ such that
\begin{equation}
N > \frac{1}{n\tilde{d}}\;\;\;, \label{a7}
\end{equation}
where $n=1,2,3,...$ and is fixed..
Then it follows that there exists at least one
pair of integers $(k_1^o,k_2^o)$ with $1 \leq k_1^o \leq N$ such
that
\begin{equation}
0 < \mu < \frac{1}{N}\;\;, \;\; \mu=\mid k_1^o\xi + k_2^o\mid \;\;\; \label{a8}
\end{equation}
(cf. \cite{kc68}).
If we denote by $m$ the integral part of $\tilde{a}/\mu$:
\begin{equation}
m=[\frac{\tilde{a}}{\mu}] \;\; \label{a9}
\end{equation}
we find
\begin{equation}
\tilde{a} \leq (m+n')\mu \leq \tilde{b}\;\;, \;\;
1 \leq n' \leq n \;\;. \label{a10}
\end{equation}
Thus our condition \ref{a3} can be always fullfilled if we choose
\begin{equation}
k_1=(m+n')k_1^o \;\;\;, \;\;\; k_2=(m+n')k_2^o \;\;\;. \label{a11}
\end{equation}
Since there is no restriction on the integer $n$ in \ref{a7}
it follows from \ref{a11} that we generate an infinite number of
pairs $(k_1,k_2)$ which satisfy our inequality \ref{a3}.

\newpage

\noindent
FIGURE CAPTIONS
\\
\\
\\
Figure 1
\\
\\
Amplitude distribution for a localized state in the
$\Phi^4$ chain (cf. text). \\
Open squares - numerical result . \\
Dashed line - predicted decay law (cf. text). \\
\\
\\
\\
Figure 2
\\
\\
Minimum of the distance of the absolute value of eigenvalue $\lambda$
from 1 as a function of the fundamental frequency $\omega_1$.
Here $|\lambda| \leq 1$, minimization is obtained with respect
to all Fourier components for a $\Phi^4$ chain with parameters
given in the text.
\\
\\
\\
Figure 3
\\
\\
Normalized amplitude entropy $\sigma_a$ versus NLE energy
for a $\Phi^4$ chain (cf. text). \\
Filled triangles - numerical result. \\
Solid line connects the triangles and serves as a guide to the eye. \\
Dashed vertical lines - predicted positions of minima of $\sigma_a$
(cf. text).
\\
\\
\\
Figure 4
\\
\\
Numerical solution for the Fourier components of the periodic
NLE of a system with $v_2=1$, $v_4=-1$, $\phi_2=0.1$, $N=100$,
$\omega_1=0.8$ (cf. text). The
absolute values of the components $A_{kl}$ are shown
as functions of the lattice site $l$ in a window of 30 lattice sites
around the NLE center. The open squares are the actual results.
The lines are guides to the eye and connect components with
same Fourier order $k$. $k$ increases from top to bottom as
$k=1,3,5,7,...,23,25$. Fourier components for even $k$ are zero
because of the symmetry of the potential.
\\
\\
\\
Figure 5
\\
\\
Slopes of the lines in Fig.4 as function of $k$ (correspond
to the exponents of the decay of the corresponding Fourier
components, cf. text) are shown as open squares. The solid
line connects the points of the theoretical prediction
using the eigenvalues of the linearized map (cf. text).
\end{document}